\documentclass[prl,superscriptaddress,twocolumn,longbibliography]{revtex4-2}
\usepackage[T1]{fontenc}
\usepackage{amsmath}
\usepackage{amssymb}
\usepackage{mathtools}
\usepackage[normalem]{ulem}
\usepackage{graphicx}
\usepackage{datetime}
\usepackage{color}
\usepackage{newtxtext,newtxmath}
\usepackage{microtype}
\usepackage{braket}
\usepackage{xr}
\usepackage{natbib}
\usepackage{hyperref}
\usepackage[utf8]{inputenc}
\usepackage{bbm}
\usepackage{tensor}
\hypersetup{
    bookmarksopen=false,
    bookmarksnumbered=true,
    pdfnewwindow=true,
    unicode=false,
    colorlinks=true,
    citecolor=blue,
    linkcolor=black,
    urlcolor=blue,
    filecolor=blue
    }
\newcommand{\vb}[1]{\boldsymbol{#1}}

\newcommand{\lsat}{\ell_\text{sat}}
\newcommand{\lvar}{\var_\tau \ell}

\DeclareMathOperator{\var}{var}

\DeclareMathOperator{\spec}{Spec}
\begin{document}
\newcommand{\figdir}{.}
\newcommand{\usal}{Departamento de F\'isica Fundamental, Universidad de Salamanca, E-37008 Salamanca, Spain}
\newcommand{\iffym}{Instituto Universitario de F\'isica Fundamental y Matem\'aticas (IUFFyM), Universidad de Salamanca, E-37008 Salamanca, Spain}

\title{Propagation of two-particle correlations across the chaotic phase for interacting bosons}

\author{\'Oscar Due\~nas}
\affiliation{\usal}
\affiliation{\iffym}
\author{David Pe\~na}
\affiliation{\usal}
\author{Alberto Rodr\'iguez}
\email[]{argon@usal.es}
\affiliation{\usal}
\affiliation{\iffym}

\begin{abstract}
We analyze the propagation of experimentally relevant two-particle correlations for one-dimensional interacting bosons, and give evidence that  many-body chaos induces the emergence of an effective diffusive regime for the fully coherent correlation dynamics, characterized by an interaction dependent diffusion coefficient, which we estimate. This result supports very recent experimental observations, and paves the way towards an efficient description of the dynamical behaviour of non-integrable complex many-body systems. Furthermore, we show that the dynamical features within experimentally accessible time scales of a conveniently defined two-particle correlation transport distance provide a direct and unambiguous characterization of many-body quantum chaos in perfect agreement with its spectral identification.
\end{abstract}

\maketitle

Complex many-body dynamics ensue from an involved interplay between many-particle interference and interactions. The latter, in particular, are responsible for the appearance of chaotic regimes characterized by universal dictates \cite{Haake2018,Borgonovi2016,Izrailev1990}, in which certain coarse-grained features of a many-body system become amenable to a description in terms of random matrix theory and statistical mechanics \cite{Deutsch1991,Rigol2008,Srednicki1996,Srednicki1999,Weidenmuller2024b}. The specificity of the system in the chaotic phase, however, remains imprinted in certain observables \cite{Pausch2020,Pausch2022,PauschThesis}, and the sensitivity to many-particle interference is in fact enhanced at the onset of such `universal' regime \cite{Schlagheck2019,Brunner2023}. This raises a fundamental issue, \emph{how does the many-particle correlation dynamical response change upon the emergence of many-body quantum chaos?}

This question bears practical implications, e.g., for the design of current quantum computing architectures \cite{Berke2020,Basilewitsch2023,Borner2023}, and its investigation  
finds an ideal experimental platform in systems of low dimensional ultracold bosons  \cite{Cheneau2012,Gring2012,Langen2013a,Trotzky2012,Ronzheimer2013a,Meinert2014a,Kaufman2016,Takasu2020, Meinert2014b,Langen2015,Rispoli2019,Lukin2018,Bohrdt2020,Leonard2023,Bordia2016,Choi2016a,Rubio-Abadal2019,Leonard2023,Wienand2023}.
For one-dimensional bosons, 
beyond the dynamics of single-particle observables \cite{Trotzky2012,Ronzheimer2013a,Meinert2014a,Kaufman2016,Takasu2020}\cite{Kollath2007,Cramer2008a,Vidmar2013,Sorg2014,Andraschko2015}, 
studies on coherent relaxation of higher-order observables from non-equilibrium configurations have mostly focused on the propagation of two-particle correlation fronts, found to form a ballistic light-cone ---for any bosonic interaction strength---, both theoretically \cite{Lauchli2008,Barmettler2012,Despres2019} and experimentally \cite{Cheneau2012}. 

Here, we show compelling evidence that the existence of many-body quantum chaos for one-dimensional bosons triggers a fundamental change in the coherent dynamics of two-particle correlations, giving rise to an effective diffusive regime governed by an interaction dependent diffusion constant. This result is in significant agreement with very recent experimental observations for hard-core bosons \cite{Wienand2023}. We find that the time development of correlations can be encoded into a suitable two-particle correlation transport distance, whose behaviour ---within experimentally reachable time scales--- further permits an explicit identification of the chaotic phase in remarkable agreement with its spectral characterization \cite{Buchleitner2003,Kolovsky2004,Biroli2010b,Kollath2010,Beugeling2014,Beugeling2015,Beugeling2015c,Dubertrand2016,Fischer2016a,Beugeling2018,DelaCruz2020,Russomanno2020,Pausch2020,Pausch2021,Pausch2022,PauschThesis}. Most interestingly, we show evidence that the equilibrium value of such correlation transport distance develops a non-analyticity (as the thermodynamic limit is approached) as a function of the interaction strength at the onset of the chaotic regime.

We model an ensemble of $N$ boson in an $L$-site one-dimensional optical lattice using the standard Bose-Hubbard hamiltonian (BHH) \cite{Lewenstein2007,Cazalilla2011,Krutitsky2016} with hard-wall boundary conditions,
\begin{equation}
 H=-J\sum_j (b^\dagger_j b_{j+1} + b_{j+1}^\dagger b_j) + \frac{U}{2}\sum_j \hat{n}_j (\hat{n}_j-1),
\end{equation}
and study the dynamics of the homogeneous Fock state  with one boson per site, $\ket{\psi_0}=\ket{1,1,\ldots,1}$, for varying relative tunneling strength $\gamma\equiv J/U>0$. The time evolution is restricted to a maximum of 200 tunneling times, $\tau\equiv Jt/\hslash \leqslant 200$, in order to work within experimentally accessible time scales, as those probed in Refs.~\cite{Rispoli2019,Leonard2023}. The chosen initial state ensures no significant net mass transport across the system in time but a $\gamma$-dependent development of many-particle correlations. We probe the dynamics of two-particle correlations via 
the experimentally accessible connected two-point density correlations, 
\begin{equation}
 C_{j,k}(\tau)\equiv\braket{\hat{n}_j(\tau) \hat{n}_k(\tau)}_{\psi_0}- \braket{\hat{n}_j(\tau)}_{\psi_0} \braket{\hat{n}_k(\tau)}_{\psi_0},
 \label{eq:c2pdc}
\end{equation}
between sites $j,k\in[1,L]$. One may conveniently define a \emph{two-particle correlation transport distance} (CTD) as 
\begin{equation}
 \ell(\tau)\equiv \sum_{d=1}^{L-1} d \braket{|C_{k,k+d}(\tau)|}_k,
 \label{eq:ctd}
\end{equation}
where angular brackets indicate an average over all site pairs $(k,k+d)$ for a given distance $d$, as was introduced and measured experimentally in Ref.~\cite{Rispoli2019}
to discern thermal
from many-body localized phases in the disordered BHH.
The CTD encodes the magnitude of all two-point density correlations in the value of a `mean' correlation distance 
\cite{SM}. As a reference, in the non-interacting limit ($\gamma\to\infty$), the asymptotic temporal average of the CTD is $\overline{\ell(\tau)}_{\gamma\to\infty} =L+O(L^{0})$ \cite{SM}, becoming equal to the system size.

\emph{Saturation regime.---}
The evolution of $\ket{\psi_0}$ for a given $\gamma$ is performed numerically via an efficient expansion of the time-evolution operator using Chebyshev polynomials \cite{Weisse2008}, which allows us to simulate accurately long time dynamics for systems up to $L=17$ at unit density (with a Hilbert space size $\mathcal{D}\simeq 10^9$) \cite{SM}. 

\begin{figure*}
  \centering
  \includegraphics[width=\textwidth]{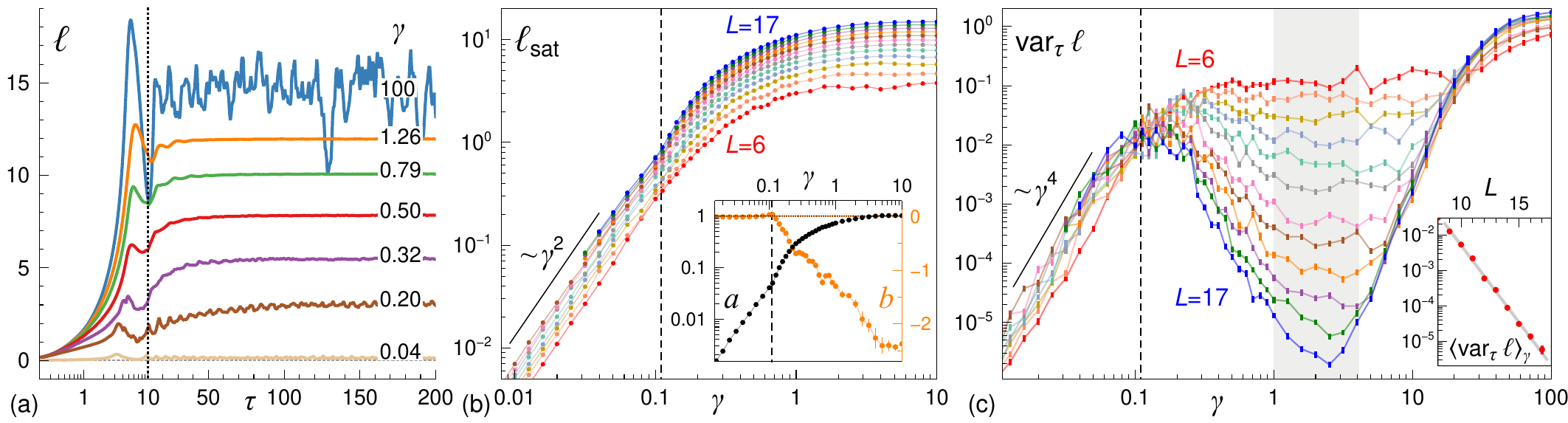}
  \caption{Dynamical features of the two-particle correlation transport distance. (a) CTD [Eq.~\eqref{eq:ctd}] versus time for $L=17$ and varying $\gamma=J/U$ as indicated. 
  Note the log (linear) scale for $\tau\leqslant 10$, time step $\Delta\tau=0.01$ ($\tau>10$, $\Delta\tau=0.5$). 
  Panels (b) and (c) show, respectively, the time averaged value $\ell_\text{sat}$ and the CTD temporal variance in the interval $\tau\in[100,200]$ versus $\gamma$ for
  $L\in[6,17]$. The vertical dashed line highlights the value $\gamma=0.11$. The inset in (b) presents the parameters of the fit $\ell_\text{sat}(\gamma)=a(\gamma)L+b(\gamma)$. 
  The inset in (c) displays $\var_\tau\ell$ averaged over the shaded $\gamma$-region for each $L$, with the solid line being the best fit $111.71 e^{-L}$. When not visible, errors are contained within symbol size.}
  \label{fig:CTDglobal}
\end{figure*}

An overview of the dynamical behaviour of $\ell(\tau)$ for $L=17$ is shown in Fig.~\ref{fig:CTDglobal}(a). The CTD undergoes an initial growth in time and, due to the finiteness of the system, eventually reaches a $\gamma$-dependent saturation value which can be well identified for $\tau\in[100,200]$.
The most interesting feature of the signals in this time interval is the striking absence of temporal fluctuations for some $\gamma$ values. The CTD saturation value, identified as the time average $\lsat\equiv \overline{\ell(\tau)}$ for $\tau\in[100,200]$, is shown in Fig.~\ref{fig:CTDglobal}(b) as a function of $\gamma$. The propagation of correlations is strongly suppressed for $\gamma<0.1$, where $\lsat\ll1$ and it exhibits a quadratic dependence on $\gamma$. As the relative tunneling strength further increases, $\lsat$ undergoes a pronounced surge, reaching a maximum followed by a subtle decay towards the non-interacting limit (see also Fig.~S2 in the supplemental material). Our analysis shows that $\lsat$ exhibits a linear growth with $L$ for all values of $\gamma$ \cite{SM}, $\lsat=aL +b$, albeit with markedly different slopes. In fact, around $\gamma=0.11$ a change in the dependence of $\lsat$ on $\gamma$ becomes visible for increasing $L$. This transition is distinctively witnessed by the fit parameters $a$ and $b$ [inset of Fig.~\ref{fig:CTDglobal}(b)], indicating that a non-analyticity develops in the vicinity of $\gamma=0.11$ in the thermodynamic limit.

The CTD fluctuations after
saturation
can be characterized by the temporal variance $\lvar\equiv \overline{\ell^2(\tau)} -\lsat^2$, which as observed in Fig.~\ref{fig:CTDglobal}(c) features three different $\gamma$-regimes. 
For low relative tunneling strength, fluctuations are weakly dependent on $L$ and strongly suppressed, growing as $\gamma^4$ up to a tipping point around $\gamma=0.11$, marking the crossover to a new regime where $\lvar$ decreases dramatically with $L$ and reaches a minimum around $\gamma\approx 3$.
Note that
$\lvar$ drops by nearly five orders of magnitude after 
a three-fold increase in $L$.
Indeed, here the fluctuations decrease exponentially with system size
[inset of Fig.~\ref{fig:CTDglobal}(c)]. This latter observation is one
implication of the eigenstate thermalization hypothesis (ETH), by which temporal fluctuations around equilibrium values are reduced exponentially with the number of degrees of freedom \cite{Srednicki1996,Srednicki1999} (i.e., the number $L$ of modes in the BHH \cite{PauschThesis}). 
Fluctuations ultimately grow again with tunneling strength, and at $\gamma\simeq 20$ a third regime appears where
$\lvar$ is slightly enhanced with system size.

\begin{figure}
  \centering
  \includegraphics[width=\columnwidth]{\figdir/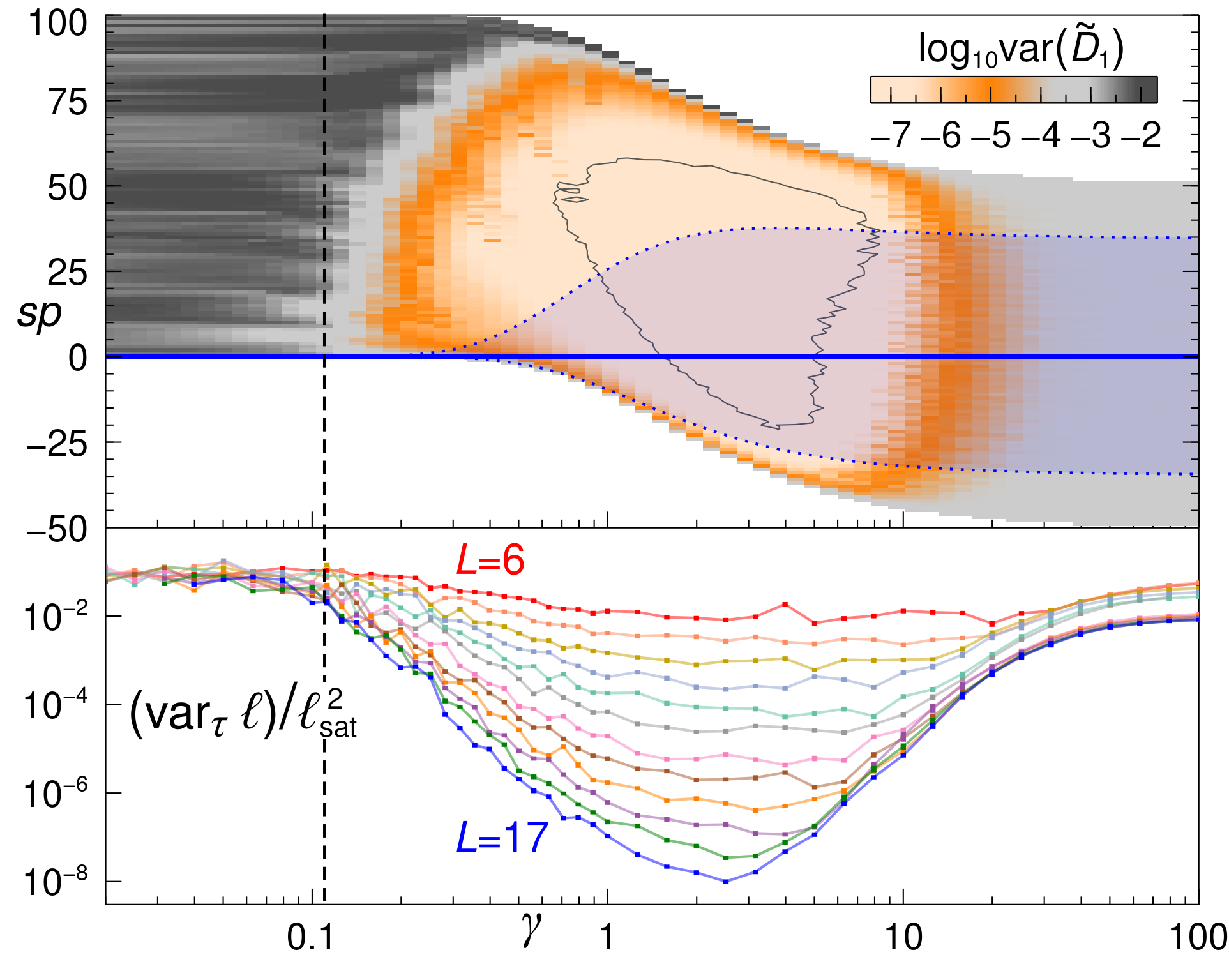}
  \caption{Comparison of spectral (top) and dynamical (bottom) pictures of the chaotic phase. The top panel shows the variance of the eigenstate fractal dimension $\tilde D_1$ [Eq.~\eqref{eq:D1}] for $L=11$ (even parity subspace, $\mathcal{D}=176\,484$) as a function of $\gamma\in[0.02,100]$ calculated for sets with 1\% of the eigenstates organized as the spectrum percentage (\textit{sp}) measured from the initial state's energy (blue line, shaded area marks the corresponding energy width) (see main text). The contour line encircles the region where $\log_{10}\var(\tilde D_1)$ lies within $5\%$ of the RMT value. The bottom panel displays the CTD relative temporal variance versus $\gamma$ for varying $L$.}
  \label{fig:CTDRevar}
\end{figure}

It is remarkable that an intermediate $\gamma$-regime of strongly suppressed temporal fluctuations in the CTD
is so clearly visible for relatively small systems within experimentally accessible time scales of $\approx100$ tunneling times.
Moreover, the relative fluctuations quantified by $(\lvar)/\lsat^2$, shown in the bottom panel of Fig.~\ref{fig:CTDRevar}, exhibit a distinctive qualitative and quantitative behaviour in this regime (which also emerges in the presence of periodic boundary conditions \cite{SM}).

The question now is to which extent this dynamical regime correlates with the system's spectrally chaotic phase. The latter is best identified from the eigenvector structure \cite{Pausch2020,Pausch2021,Pausch2022,PauschThesis}, which is efficiently probed by the generalized fractal dimension
\begin{equation}
 \tilde{D}_1 = -\sum_{\ket{\vb{n}}} |\braket{\vb{n}|\psi}|^2 \log |\braket{\vb{n}|\psi}|^2 /\log \mathcal{D},
 \label{eq:D1}
\end{equation}
for eigenstates $\ket{\psi}$ expanded in the on-site Fock basis $\{\ket{\vb{n}}\}$. The top panel of Fig.~\ref{fig:CTDRevar} shows the energy resolved variance of $\tilde{D}_1$ for $L=11$ calculated over sets with 1\% of the eigenstates, as a function of $\gamma$ and of the spectral percentage distance (\textit{sp}) from the energy $E=0$ of the initial state $\ket{\psi_0}$ taken as a reference value (i.e., the spectral percentage at energy $E$ corresponds to $sp=100\times[\mathcal{N}(E)-\mathcal{N}(0)]/\mathcal{D}$, with $\mathcal{N}(E)$ the integrated density of states).

In this stationary picture, as demonstrated in Refs.~\cite{Pausch2020,Pausch2021,Pausch2022,PauschThesis}, the chaotic phase is revealed by the region with suppressed $\var(\tilde{D}_1)$, which correlates with the emergence of eigenstates that become extended ($\tilde{D}_1\to 1$) in Fock space in the thermodynamic limit \cite{SM}. We note that the onset of the chaotic domain so identified does not change with increasing $L$ \cite{SM}. To see how the initial state participates of the system eigenstates, the reference energy trajectory at \textit{sp} $=0$ is flanked by two bands corresponding to $\pm \sigma \equiv \pm (\braket{H^2}_{\psi_0}-\braket{H}^2_{\psi_0})^{1/2}$, which gives the energy width of the local density of states in Fock space at the point $\ket{\psi_0}$. The top panel of Fig.~\ref{fig:CTDRevar} thus shows how the spectrally chaotic phase develops as a function of $\gamma$ from the perspective of $\ket{\psi_0}$, whose passage through the chaotic phase (accompanied by a marked broadening of the energy width) correlates unambiguously with the regime of suppressed temporal fluctuations of the CTD.
Furthermore, $(\lvar)/\lsat^2$ is minimized in the $\gamma$-range where the initial state participates maximally of eigenstates that exhibit the closest agreement with
random matrix theory (RMT) (see contour line in the top panel of Fig.~\ref{fig:CTDRevar}). These results also confirm that the dynamical emergence of ETH hinges on the existence of chaotic (extended in Fock space) eigenstates.

\emph{Pre-saturation dynamics.---}
One should also ask if and how the emergence of the chaotic phase would manifest itself dynamically through the CTD before saturation. This requires challenging simulations
for large enough systems, exhibiting a regime of steady growth for $\ell(\tau)$ sufficiently far away from the pronounced maxima that precede the onset of saturation [cf.~Fig.~\ref{fig:CTDglobal}(a)].

Let us first understand the CTD limiting behaviours.
For short times,
an exact calculation
up to order $\tau^4$ yields \cite{SM}, 
\begin{equation}
 \ell(\tau)\underset{L\gg1}{=}4\tau^2 - \left(6 +\frac{1}{3\gamma^2}\right)\tau^4 + O(\tau^6).
 \label{eq:elltau2}
\end{equation}
Hence, the CTD exhibits an initial quadratic growth whose extension is maximal in the non-interacting case ($\gamma\to\infty$) and shrinks as $\sim\gamma$ for strong interactions $(\gamma\to0^+$).

Asymptotically, for $L\gg\tau\to\infty$, the analysis in the absence of interactions reveals \cite{SM}
\begin{equation}
 \ell(\tau) \underset{\gamma\to\infty}{=}\frac{16}{\pi^2}\tau+O\left(\frac{\log{\tau}}{\tau}\right),
 \label{eq:CTDgammainfasym}
\end{equation}
and hence ballistic spreading.
In the opposite limit of very strong interaction ($\gamma\ll 1$), making use of the fermionization approach of Ref.~\cite{Barmettler2012}, we compute an analytical expression for the CTD for $L=\infty$ \cite{SM}. Most importantly, this expression provides the exact temporal asymptotic behaviour in the very limit $\gamma\to0^+$,
\begin{equation}
 \ell(\tau) \underset{\gamma\to0^+}{=} \frac{64\gamma^2}{\pi} \tau+O(\tau^{-1}).
 \label{eq:CTDgamma0asym}
\end{equation}
Therefore, the spreading of two-point density correlations is also ballistic when approaching the $\gamma=0$ integrable point. 
It is precisely in the dependence of the CTD's steady growth on $\gamma$ where the emergence of the chaotic phase may be observable. 
(Recall that
$\ell_\text{sat}$ depends linearly on $L$,
and hence the CTD's increase as $L\gg\tau\to\infty$ is unbounded for all $\gamma$.)

\begin{figure}
  \centering
  \includegraphics[width=\columnwidth]{\figdir/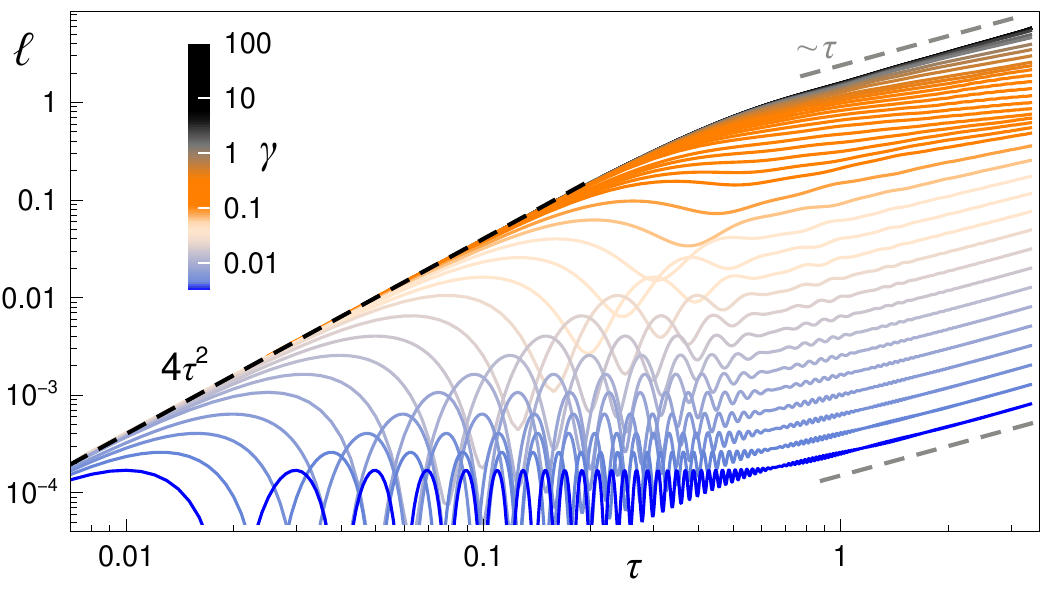}
  \caption{Time evolution of the CTD for $L=100$ and varying $\gamma\in[0.00316,100]$
  as indicated by the color scale. The black dashed line corresponds to the initial quadratic growth $4\tau^2$ [cf.~Eq.~\eqref{eq:elltau2}], while gray dashed lines highlight ballistic behaviour.}
  \label{fig:CTD-mps}
\end{figure}

In Fig.~\ref{fig:CTD-mps}, we show the behaviour of the CTD in time for $L=100$ and varying $\gamma$, obtained using time evolving block decimation for matrix product states.
The simulation parameters are carefully chosen to ensure convergence of the signals with a tolerance $\lesssim0.5\%$ \cite{SM}.
As can be observed, the initial growth is dictated by Eq.~\eqref{eq:elltau2}, where the extent of the quadratic regime wanes with the relative interaction strength. For dominating interactions ($\gamma\ll1$), the departure from the $\tau^2$ behaviour is followed by an oscillatory regime governed by
transitions to doublon and holon states, characterized by the single frequency $\omega=\gamma^{-1}$. As time progresses, the doublon-holon propagation and the participation of further states slowly builds up and eventually the CTD exhibits a sustained enhancement in time, even if with a very small amplitude. 
(This behaviour can be analytically described in the limit $\gamma\to0^+$ \cite{SM}.) For weaker interactions, the oscillatory regime is progressively reduced and the quadratic time dependence transitions smoothly into a different behaviour. 

The relevant steady growth of $\ell(\tau)$, after the uneventful $\tau^2$ and oscillatory evolutions, is visible in Fig.~\ref{fig:CTD-mps} for $\tau>1$ and for all $\gamma$, where
the CTD seems to converge to a guiding power-law tendency. Note that, despite the limited simulation times, the predicted asymptotic ballistic behaviour in the limits $\gamma\to\infty$ and $\gamma\to0^+$ is  unambiguously observable.
For intermediate $\gamma$, however, the CTD's growth
registers a noticeable slowdown.  

\begin{figure}
  \centering
  \includegraphics[width=\columnwidth]{\figdir/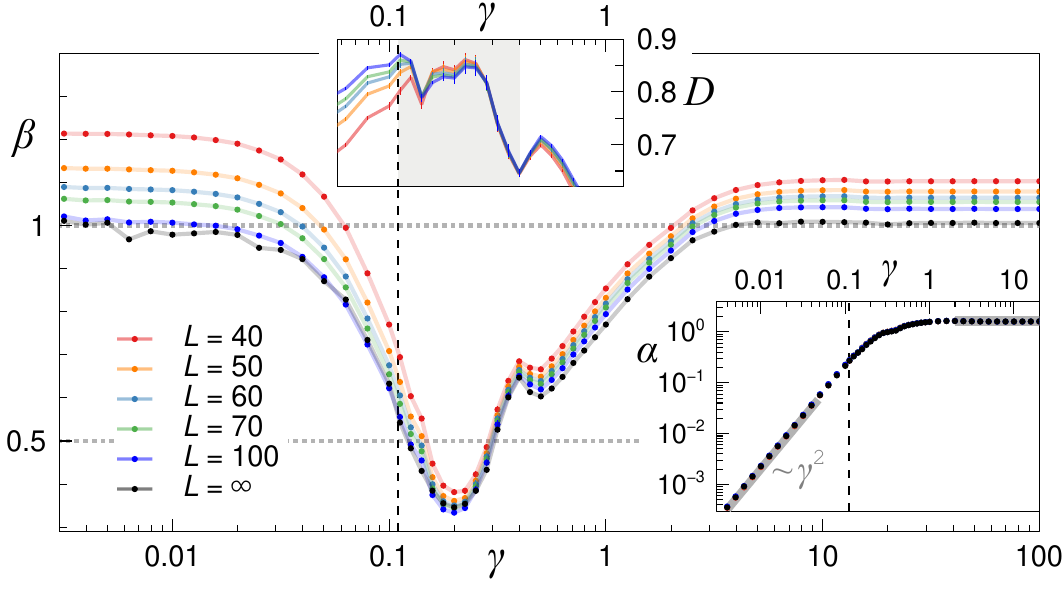}
  \caption{Parameters $\beta$ (main panel) and $\alpha$ (lower inset) of the CTD fit $\ell(\tau)=\alpha\tau^\beta$ in the interval $\tau\in[2.2,3.3]$, and diffusion constant $D$ (upper inset) as functions of $\gamma$ for the indicated system sizes.
  Horizontal dotted lines highlight ballistic ($\beta=1$) and diffusive ($\beta=1/2$) behaviour,
  while the vertical dashed line marks the value $\gamma=0.11$. In the lower inset, all data for different $L$ collapse, and the thick gray lines
  correspond to the coefficients of Eqs.~\eqref{eq:CTDgammainfasym} and \eqref{eq:CTDgamma0asym}. 
  When not visible, errors are contained within symbol size.}
  \label{fig:CTD-PLfits}
\end{figure}

In order to quantify the steady increase of $\ell(\tau)$, we show in Fig.~\ref{fig:CTD-PLfits} the parameters of the fit $\ell(\tau)=\alpha \tau^\beta$ for $\tau\in[2.2,3.3]$ as functions of $\gamma$ for $L\in[40,100]$. In this time interval, the CTD shows a convergence towards a system-size independent form \cite{SM}, allowing us to estimate the signal for $L=\infty$, and their corresponding fit parameters, included in Fig.~\ref{fig:CTD-PLfits}.
As revealed by the value of the exponent $\beta$, the growth of the CTD is indeed ballistic near the integrable limits $\gamma\to 0^+$ and $\gamma\to\infty$.
As the relative tunneling strength departs from the region $\gamma\ll 1$, the value of $\beta$ shows initially a timid decrease that eventually evolves into a marked drop reaching the value 1/2 around $\gamma=0.11$, the threshold identified in our previous analyses. Beyond this point, the exponent $\beta$ seems to want 
to develop an oscillation around 1/2, which is suddenly interrupted at $\gamma\approx 0.4$. For larger $\gamma$, the exponent approaches monotonically 1, and ballistic growth is fully restored for $\gamma\gtrsim 3$. 
It is worth emphasizing that the
fit amplitude, $\alpha$ (lower inset to Fig.~\ref{fig:CTD-PLfits}), is well converged with $L$, and its dependence on $\gamma$ is consistent with the behaviour observed for $\ell_\text{sat}$ (and its fit parameter $a$) [cf.~Fig.~\ref{fig:CTDglobal}(b)], as well as with Eq.~\eqref{eq:CTDgammainfasym} for $\gamma\to\infty$ and Eq.~\eqref{eq:CTDgamma0asym} for $\gamma\to0^+$.

The results indicate that the onset of the chaotic phase correlates with the emergence of a diffusive behaviour for the CTD. The spreading of two-particle correlations in the chaotic regime would then be governed by a diffusive process characterized by a single parameter, the
diffusion constant $D(\gamma)$.
Diffusion seems plausible in the range $0.1\lesssim\gamma\lesssim 0.4$, that, however, does not correlate positively with the entire chaotic phase as identified previously. 
This (as well as the fluctuation of $\beta$ around $1/2$) may be attributed to the short time scales available:
If the characteristic mean free time of the diffusive process exceeded the extent of the considered time interval then ballistic dynamics should be effectively realized. 
Accordingly, the observable diffusive regime is primarily limited by the accessible time scales, rather than by just the extension of the chaotic phase.

A comparison of the CTD against
a simple diffusive model \cite{SM} permits a qualitative estimation of the diffusion constant, shown in the upper inset of Fig.~\ref{fig:CTD-PLfits}. Upon entering the chaotic phase ($\gamma\approx 0.11$), the estimated $D$ quickly collapses for all system sizes, and exhibits a slight decay upon approaching $\gamma\approx 0.4$, which is arguably the end of the observable diffusive regime. We note that the diffusion constant lingers around the value $D\approx 0.8\, Ja^2/\hbar$ ($a$ being the lattice constant), in agreement with a very recent experimental observation for hard-core bosons on a ladder \cite{Wienand2023}. 

The identified diffusive $\gamma$-range for two-particle correlations may be compared
with the results for one-particle observables of Ref.~\cite{Andraschko2015}, where diffusive mass transport of a single density defect was found for $\gamma\in[0.2,0.5]$ on similar time scales, as well as with the pronounced slowdown in the expansion of a 1D bosonic cloud experimentally observed for $\gamma\approx0.3$ in Ref.~\cite{Ronzheimer2013a}, conjectured to be due to the emergence of diffusive dynamics. 

Note that diffusive spreading of correlations in the many-body chaotic phase lies beyond the reach of standard RMT ensembles. For the Gaussian Orthogonal Ensemble (GOE), one may evaluate $C_{j,k}^{\textrm{GOE}}(\tilde\tau)$ analytically, with time $\tilde\tau$ in units of the inverse of the spectral width (results will be reported elsewhere). Here, correlations are independent of the distance $|j-k|$ and spread as $\sim \tilde\tau^2$ before saturation. Hence, the sparse Fock-space connectivity of typical many-body systems seems essential for the emergence of diffusive dynamics.

We have studied the propagation of experimentally relevant two-particle correlations for one-dimensional interacting bosons. Our results show evidence that in the chaotic regime certain aspects of the otherwise complex many-particle fully coherent dynamics are amenable to a simplified approach using a diffusive model specified by an interaction dependent diffusion coefficient. This observation goes in hand with very recent experimental results for low-dimensional hard-core bosons \cite{Wienand2023}, and paves the way towards an efficient description of the dynamical behaviour of non-integrable complex many-body systems. Furthermore, the time development of two-point density correlations can be conveniently encoded in a two-particle correlation transport distance (CTD), whose features within experimentally accessible time scales provide a direct and unambiguous characterization of many-body quantum chaos. Besides demonstrating analytically the ballistic growth of the CTD in the vicinity of the system's integrable limits, we specifically show that \emph{(i)} the onset of many-body chaos correlates with the emergence of a non-analyticity in the equilibrium value of the CTD when approaching the thermodynamic limit, and \emph{(ii)} the regime of suppressed CTD temporal fluctuations expose the chaotic phase in a one-to-one correspondence with its spectral identification.

Given that the system's dynamical response is fundamentally shaped by many-particle interference \cite{Tichy2012a,Engl2014,Menssen2017,Brunner2018,Dittel2018,Giordani2018,Rammensee2018,Tomsovic2018,Dufour2020,Walschaers2020}, controlled by particle distinguishability, most prominently at the onset of chaos \cite{Schlagheck2019,Brunner2023},
an analysis of the CTD and the diffusion constant for bosonic mixtures could reveal how particle distinguishability may enter into effective macroscopic descriptions of many-body chaotic dynamics.

\begin{acknowledgments}
We thank Lukas Pausch, Edoardo Carnio and Andreas Buchleitner for helpful discussions. The authors acknowledge support by Spanish MCIN/AEI/10.13039/501100011033 through Grant No.~PID2020-114830GB-I00. A.R.~acknowledges support by the German Research Foundation (DFG) through Grant No.~402552777. O.D.~acknowledges support from a Ph.D Fellowship funded by Consejer\'ia de Educaci\'on de la Junta de Castilla y Le\'on and European Social Fund Plus. This research has made use of the high performance computing resources of the Castilla y Le\'on Supercomputing Center (SCAYLE, www.scayle.es), financed by the European Regional Development Fund (ERDF), and of the CSUC (Consorci de Serveis Universitaris de Catalunya) supercomputing resources. We thankfully acknowledge RES resources provided by the Galician Supercomputing Center (CESGA) in FinisTerrae III to activity FI-2024-2-0027. The supercomputer FinisTerrae III and its permanent data storage system have been funded by the Spanish Ministry of Science and Innovation, the Galician Government and the European Regional Development Fund (ERDF).
\end{acknowledgments}

\section{Supplemental Material}

\section{Analytical results for the correlation transport distance}
\subsection{Definition}
The two-particle correlation transport distance (CTD) [see Eq.~\eqref{M-eq:ctd} in main manuscript] is defined as
\begin{equation}
 \ell(\tau)=\sum_{d\geqslant0} d\,P(d,\tau),
 \label{eq:ctdSM}
\end{equation}
where
\begin{equation}
 P(d,\tau)\equiv \braket{|C_{k,k+d}(\tau)|}_k
 \label{eq:Pd}
\end{equation}
is the average connected two-point density correlation at distance $d$ [see Eq.~\eqref{M-eq:c2pdc}]. The CTD can then be seen as the first moment of the unnormalized discrete distribution $P(d,\tau)$ for the distance, that we denote by $\langle d \rangle=\ell(\tau)$. As shown in Fig.~\ref{fig:normP}, the norm $N(\tau)\equiv\langle 1 \rangle$ of $P(d,\tau)$ saturates in time, and hence the alternative definition of a normalized CTD  $\ell'(\tau)\equiv \ell(\tau)/N(\tau)$ exhibits the same power-law exponents governing the asymptotic temporal growth, as we have checked. However, such normalized CTD would not carry the information on the magnitude of two-particle correlations, and hence definition $\eqref{eq:ctdSM}$ seems preferable in most cases.

\subsection{Saturation value in the non-interacting limit}

In the \emph{non-interacting limit} ($\gamma=J/U \to\infty$), the Heisenberg representation of the field operators in the Wannier basis is exactly accessible. For periodic boundary conditions (PBC),
\begin{equation}
 b_j(\tau)=\sum_{s=1}^L F_{js}(\tau) b_s, \quad F_{js}(\tau)\equiv \frac{1}{L}\sum_{k=1}^L e^{i(j-s)\phi(k)}e^{i2\tau\cos\phi(k)},
 \label{eq:Fjs}
\end{equation}
with $\phi(k)=2\pi k/L$. The connected two-point density correlations [see Eq.~\eqref{M-eq:c2pdc}] for the initial Fock state $\ket{\psi_0}=\ket{1,1,\ldots,1}$ are given by
\begin{equation}
 C_{|j-j'|}(\tau) = 2\delta_{j,j'}-2\sum_{r=1}^L |F_{jr}(\tau)F_{j'r}(\tau) |^2,
 \label{eq:Cjj}
\end{equation}
and are only dependent upon the distance $|j-j'|$. Therefore
\begin{equation}
 P(d,\tau)\equiv 2 \sum_{r=-d_\text{max}}^{d_\text{max}-\eta} |F_{0r}(\tau)F_{dr}(\tau) |^2, \quad 1\leqslant d\leqslant d_\text{max},
 \label{eq:PdNi}
\end{equation}
after using the cyclic freedom of the indices and with $d_\text{max}=L/2$ [$d_\text{max}=(L-1)/2$] and $\eta=1$ [$\eta=0$] for even [odd] $L$.

The asymptotic time average of \eqref{eq:Cjj} for fixed $L$ can be performed analytically, and one finds
\begin{equation}
 \overline{|C_d(\tau)|}=2\times
 \begin{cases} \displaystyle
  \frac{1}{L}-\frac{1}{L^2}, & \!\!\text{odd $L$}, \\ \displaystyle
  \frac{1}{L}- \frac{2+\delta_{d,L/2}(-1)^{L/2}}{L^2} + \frac{1+(-1)^{L/2}}{L^3}, &  \!\!\!\text{even $L$},
 \end{cases}
\end{equation}
for $1\leqslant d\leqslant d_\text{max}$. This permits to obtain the saturation value of the correlation transport distance (CTD),
\begin{equation}
 \overline{\ell(\tau)}_\text{PBC}\underset{\gamma\to\infty}{=}\sum_{d=1}^{d_\text{max}} d \overline{|C_d(\tau)|} = \frac{L}{4} + O(L^0).
\end{equation}
For hard-wall boundary conditions (HWBC), an expression for the two-point density correlations can similarly be found, but the calculation of the time average is more involved. Nonetheless, for large $L$, correlations involving inner sites (not edges) should become determined only by the distance $d$ and the dominant term in $\overline{|C_d(\tau)|}$ should be the same as for PBC. Hence, the behaviour of the CTD saturation value may be inferred to be
\begin{equation}
 \overline{\ell(\tau)}\underset{\gamma\to\infty}{=}\sum_{d=1}^{L-1} d \left[2/L+O(L^{-2})\right] = L + O(L^0),
\end{equation}
as can also be numerically confirmed.

\subsection{Short time behaviour}

The exact calculation of non-vanishing correlations up to order $\tau^4$ in the interacting case for HWBC yields
\begin{align}
 C_{j,j+1}(\tau)&=-4\tau^2\!+\!\left[12\!-\!\frac{10}{3}(\delta_{j,1}+\delta_{j+1,L})+\frac{1}{3\gamma^2}\right]\tau^4+O(\tau^6), \\
 C_{j,j+2}(\tau)&=-3\tau^4+O(\tau^6),
\end{align}
which are converged for sizes $L\geqslant 4$. We note that the dominant contribution to the correlation short time expansion is always interaction independent. The short time behaviour of the CTD is then
\begin{equation}\label{eq:CTD-Short-Time}
 \ell(\tau) = 4\tau^2 - \left(6 -\frac{20}{3(L-1)}+\frac{1}{3\gamma^2}\right)\tau^4 + O(\tau^6).
\end{equation}
From the comparison of the first two terms, one sees that the initial quadratic growth in time should strictly hold for $\tau\ll \tau_*\equiv\gamma\sqrt{12/(18\gamma^2+1)}$ (for $L\gg 1$), i.e., the quadratic regime is maximal in the noninteracting limit ($\tau_*\approx0.82$) and shrinks $\sim\gamma$ for strong interactions ($\gamma\to0^+$).

\subsection{Long time behaviour}
In the \emph{non-interacting case}, for $L\to\infty$ the distribution \eqref{eq:PdNi} can formally be written using Bessel functions as
\begin{equation}
 P(d,\tau) = 2 \sum_{r\in\mathbb{Z}} \left|J_r(2\tau)J_{d-r}(2\tau) \right|^2, \quad d\geqslant 1.
\end{equation}
Expressing the terms in the series as Fourier coefficients \cite{Martin2008}, one arrives at the following integral representation,
\begin{equation}
 P(d,\tau) = \frac{1}{\pi}\int_{-\pi}^{\pi} J_0^2(4\tau\sin[\theta/2]) e^{i\theta d} d\theta, \quad d\geqslant 1,
\end{equation}
which allows for the exact computation of several `moments' of $P(d,\tau)$. We find \cite{Stoyanov1987}
\begin{align}
 N(\tau) &= \sum_{d\geqslant 0} P(d,\tau) = 3\left[1 - \tensor[_2]{F}{_3}(\tfrac{1}{2},\tfrac{1}{2};1,1,1;-16\tau^2)\right] \notag\\
 &= 3-\frac{3}{2\pi^2\tau} (\log\tau+\gamma+6\log 2) +O(\tau^{-3/2}), \label{eq:normP} \\
 \langle d \rangle &= \sum_{d\geqslant1} d\, P(d,\tau) = 4\tau^2 \tensor[_2]{F}{_3}(\tfrac{1}{2},\tfrac{3}{2};2,2,2;-16\tau^2) \notag\\
 &= \frac{16}{\pi^2}\tau-\frac{\log\tau+6\log 2+\gamma-1/2}{8\pi^2\tau} +O(\tau^{-3/2}),
\end{align}
where $\gamma$ is Euler's constant, and $\tensor[_p]{F}{_q}$ denotes the hypergeometric function. Therefore, for $L\to\infty$ in the non-interacting case as $\tau\to\infty$ the norm $N(\tau)$ of the distribution $P(d,\tau)$ converges to a steady value, and the asymptotic growth of the CTD for $\ket{\psi_0}$ is ballistic.

In the \emph{strongly interacting limit} ($\gamma\to 0^+$), the CTD for $L=\infty$ also admits a closed analytical expression. Using the fermionization approach developed in Ref.~\cite{Barmettler2012} to describe the dynamics in this limit in terms of propagating doublons and holons, one can write
\begin{equation}
C_d(\tau)=-\sum_{\sigma}\left( \left\lvert g_d^{\sigma,\sigma}(\tau) \right\rvert^2+\left\lvert g_d^{\sigma,\bar{\sigma}}(\tau) \right\rvert^2 \right),
\label{eq:corr_gamm0}
\end{equation}
where $\sigma = +, -,$ indicates the quasiparticle type ($+$ corresponding to doublons and $-$ to holons), $\bar{\sigma}=-\sigma$ and
\begin{align}
\notag g_d^{\sigma,\sigma}(\tau) &=\left\langle c^\dagger_{j+d,\sigma}(\tau) c_{j,\sigma}(\tau) \right\rangle = \\
&= \frac{1}{2\pi}\int^{\pi}_{-\pi}d k e^{-i k d}\left\langle c^\dagger_{k,\sigma}(\tau)c_{k,\sigma}(\tau)\right\rangle, \\
\notag g_d^{\sigma,\bar{\sigma}}(\tau) &= \left\langle c_{j+d,\sigma}(\tau) c_{j,\bar{\sigma}}(\tau) \right\rangle = \left\langle c^\dagger_{j,\sigma}(\tau) c^\dagger_{j+d,\bar{\sigma}}(\tau) \right\rangle^*= \\
&= \frac{1}{2\pi}\int^{\pi}_{-\pi}d k e^{-i k d}\left\langle c^\dagger_{k,\sigma}(\tau)c_{-k,\bar{\sigma}}(\tau)\right\rangle,
\end{align}
with $c^\dagger_{j,\sigma}, c_{j,\sigma}$ the fermionic creation and annihilation operators of quaisparticle $\sigma$ at site $j$, and $c^\dagger_{k,\sigma}, c_{k,\sigma}$ the corresponding operators for quasiparticle $\sigma$ with momentum $k$. Now, in this limit and for our initial state (we drop the explicit dependence on $\tau$ of the operators for clarity)
\begin{align}
\left\langle c_{k,\sigma}c_{-k,\bar{\sigma}} \right\rangle &= i2\sqrt{2}\gamma\sin(k)\left[ e^{-i2\omega(k)\tau}-1 \right]+O\left( \gamma^3 \right), \\
\left\langle c^\dagger_{k,\sigma}c_{k,\sigma} \right\rangle &= -16\gamma^2\sin^2(k)\left\lbrace \cos[2\omega(k)\tau]-1 \right\rbrace+O\left( \gamma^4 \right),
\end{align}
where $\omega(k) = 1/(2\gamma)-3\cos(k)+O(\gamma)$. These expressions (notice the independence from the specific value of $\sigma$) imply that the term  $\left\lvert g^{\sigma,\sigma}_d(\tau) \right\rvert^2\sim O\left( \gamma^4 \right)$, and can be neglected. On the other hand, for the only relevant term one has
\begin{align}
\left\lvert g^{\sigma,\bar{\sigma}}_1(\tau) \right\rvert &\underset{\gamma\to 0^+}{=} 2\gamma^2\left\lbrace 1-\frac{2J_1(6\tau)}{3\tau}\cos\left( \frac{\tau}{\gamma} \right)+\frac{J_1^2(6\tau)}{9\tau^2} \right\rbrace,\\
\left\lvert g^{\sigma,\bar{\sigma}}_d(\tau) \right\rvert &\underset{\gamma\to 0^+}{=} \frac{2d^2\gamma^2}{9}\frac{J^2_d(6\tau)}{\tau^2}, \quad d > 1.
\end{align}
Inserting these into equation \eqref{eq:corr_gamm0}  and using the definition of the CTD, one finally arrives to
\begin{align}
 \ell(\tau)\underset{\gamma\to 0^+}{=} 4\gamma^2\bigg\{ & 1+ (1+48\tau^2)J_0^2(6\tau) -8\tau J_0(6\tau)J_1(6\tau) \notag\\
 & +\left(\frac{1}{3}+48\tau^2\right)J_1^2(6\tau)
 - \frac{J_1(6\tau)}{3\tau}2\cos(\tau/\gamma) \bigg\},
 \label{eq:CTDgamma0}
\end{align}
which provides a good description of the numerical data in this regime, as we show later on (Fig.~\ref{fig:CTD-analyticalVSnumerics}). Most importantly, this expression describes exactly the asymptotic dynamics ($\tau\to\infty$) in the very limit $\gamma\to 0^+$. The asymptotic expansion of Eq.~\eqref{eq:CTDgamma0} yields
\begin{equation}
\ell(\tau) \underset{\gamma\to 0^+}{=} \frac{4\gamma^2}{\pi}\left[16\tau + \frac{1}{6\tau}+O(\tau^{-3/2})\right].
\end{equation}
One may also calculate the asymptotic expansion of the norm $N(\tau)$ in this limit
\begin{equation}
 N(\tau) \underset{\gamma\to 0^+}{=} 24\gamma^2+O(\tau^{-3/2}).
 \label{eq:normgamma0}
\end{equation}
Therefore, one arrives at the same conclusions as in the non-interacting case: The norm $N(\tau)$ of the distribution $P(d,\tau)$ converges to a steady value and the asymptotic growth of the CTD for $\ket{\psi_0}$ is ballistic.

\section{Numerical simulation of $\boldsymbol{\ell(\tau)}$ for long times}
Given the Bose-Hubbard hamiltonian, the time evolution of a generic initial state $\ket{\psi_0}$ can be numerically implemented by expanding the time evolution operator $\mathcal{U}(\tau)=e^{-i(H/J)\tau}$, with $\tau\equiv Jt/\hslash$, in terms of Chebyshev polynomials of the first kind $T_n(x)$, $n\geqslant 0$ \cite{Weisse2008}. First, a rescaled hamiltonian $\tilde{H}$ obeying $\spec(\tilde{H})\in[-1,1]$ must be defined as $H/J=a\tilde{H}+b\mathbbm{1}$, with $a=(E_\text{max}-E_\text{min})/2J$, $b=(E_\text{max}+E_\text{min})/2J$. The forward propagation of a state from time $\tau$ to $\tau+\Delta\tau$ can be written as
\begin{equation}
 \mathcal{U}(\Delta\tau)\ket{\psi(\tau)}\simeq e^{-ib\Delta\tau}\left[c_0\ket{v_0(\tau)} + 2\sum_{n=1}^M c_n \ket{v_n(\tau)} \right],
 \label{eq:Uexp}
\end{equation}
where $c_n\equiv (-i)^n J_n(a\Delta\tau)$, $J_n(x)$ being the Bessel functions, $M$ is the chosen cut-off for the expansion, and the involved vector states are retrieved from the recursion relation of the Chebyshev polynomials,
\begin{subequations}
\begin{align}
 \ket{v_0(\tau)} &\equiv \ket{\psi(\tau)}, \\
 \ket{v_1(\tau)} & \equiv T_1(\tilde{H})\ket{v_0(\tau)} =\tilde{H} \ket{\psi(\tau)}, \\
 \ket{v_{n+1}(\tau)} &\equiv T_{n+1}(\tilde{H})\ket{v_0(\tau)} = 2\tilde{H}\ket{v_n(\tau)}- \ket{v_{n-1}(\tau)}.
\end{align}
\label{eq:recurV}
\end{subequations}
This technique is numerically efficient for two reasons. Firstly, for a given value of $a\Delta\tau$, the coefficients $c_n$ decay super exponentially with $n$ for large $n$, and hence an accurate expansion may be achieved with a reasonable number of terms, although for a desired precision, the larger the spectral width $a$ and/or the chosen time step $\Delta\tau$, the larger the number of required terms. Secondly, the implementation of $\eqref{eq:Uexp}$, according to Eqs.~\eqref{eq:recurV}, only requires matrix-vector multiplications, which can be efficiently parallelized \cite{petsc-user-ref,petsc-efficient,petsc-web-page,slepc}.

\begin{figure}
  \centering
  \includegraphics[width=.9\columnwidth]{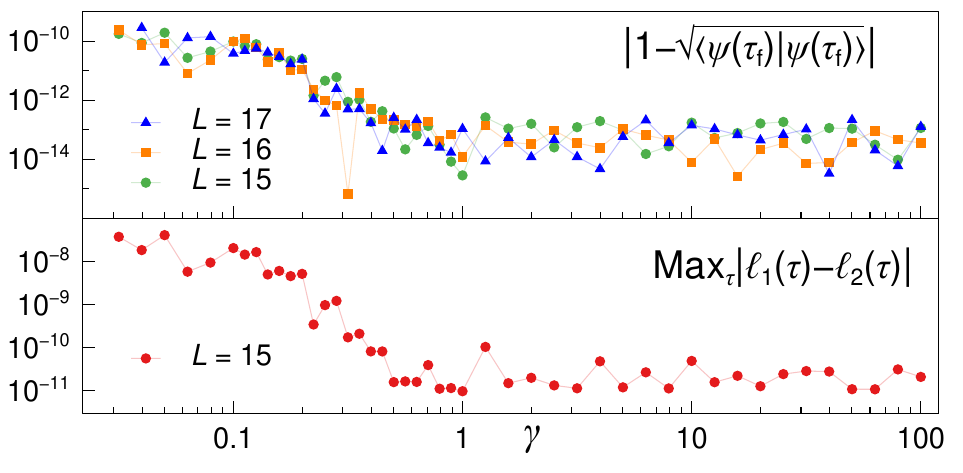}
  \caption{Accuracy analysis of time evolution as a function of $\gamma= J/U$. The top panel shows the error in the norm of the state at the final simulation time $\tau_f=200$ with the cut-off in the expansion coefficients $|c_n|$ at $10^{-12}$ for different system sizes. The bottom panel displays the maximum difference in the correlation transport distance for any time $\tau\in[0.5,200]$ between the results with cut-offs at $10^{-12}$ [$\ell_1(\tau)$] and at $10^{-16}$ [$\ell_2(\tau)$] for $L=15$.}
  \label{fig:Errors}
\end{figure}

Our implementation of this technique does not involve any truncation of the maximum occupation number in the Fock basis, and since the initial state exhibits even parity symmetry, the dynamics need only be carried out in the corresponding irreducible subspace. The expansion cut-off $M$ is chosen to keep all terms with coefficients obeying $|c_n|\geqslant 10^{-12}$ for any $\gamma=J/U$. The error induced by the truncation grows linearly with the number of time steps with a $\gamma$-sensitive slope. We have checked that the chosen precision is more than enough to ensure convergence of the calculated density correlations and of $\ell(\tau)$, as demonstrated in Fig.~\ref{fig:Errors}.

\begin{figure}
  \centering
  \includegraphics[width=.95\columnwidth]{\figdir/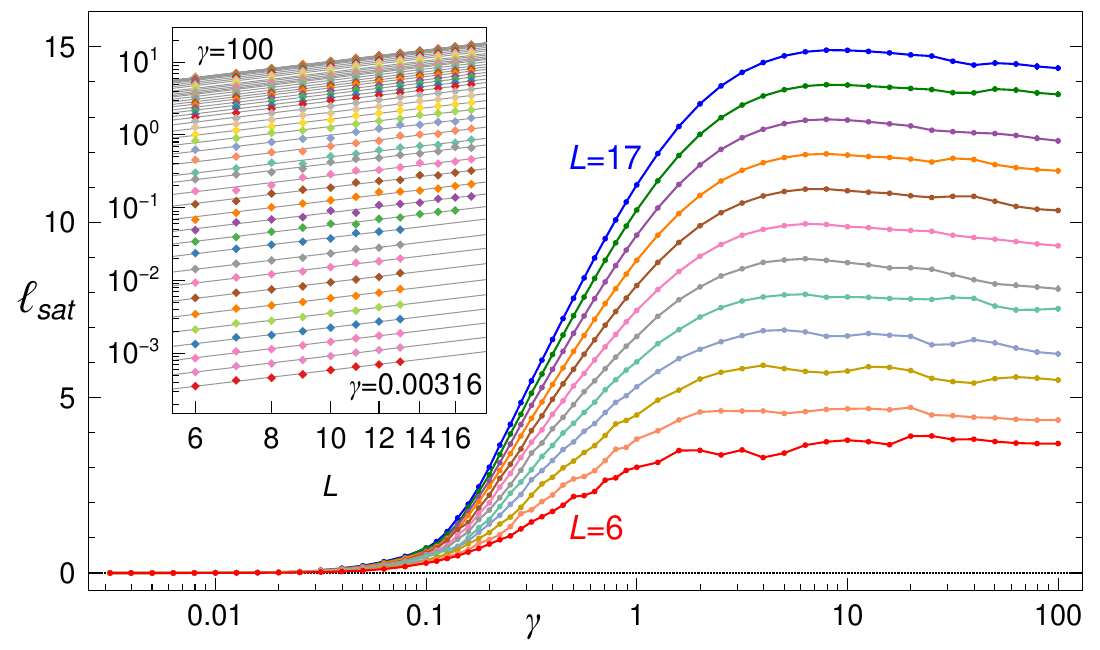}
  \caption{CTD time averaged value for $\tau\in[100,200]$ versus $\gamma$ for system sizes $L=6$ to $L=17$ from bottom to top. The inset shows the data (symbols) after the fit $\ell_\text{sat}(\gamma)=a(\gamma)L+b(\gamma)$ (solid lines), in the form $\ell_\text{sat}-b$ versus $L$ in log-log scale for 56 values of $\gamma$ from $\gamma=0.00316$ (bottom) to $\gamma=100$ (top). When not visible, errors are contained within symbols size.}
  \label{fig:CTDsat}
\end{figure}
The time averaged value of the CTD, $\ell_\text{sat}$, in the range $\tau\in[100,200]$, displayed in Fig.~\ref{M-fig:CTDglobal}(b) of the manuscript is represented in linear scale in Fig.~\ref{fig:CTDsat}. The inset demonstrates how $\ell_\text{sat}$ increases linearly with system size for any value of $\gamma$, and the data are accompanied by the corresponding linear fits.

\begin{figure}
  \centering
  \includegraphics[width=.9\columnwidth]{\figdir/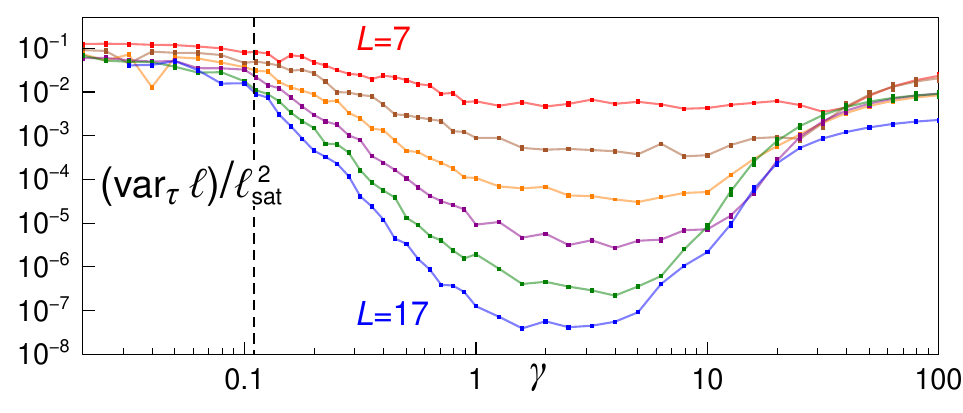}
  \caption{CTD relative temporal variance in $\tau\in[100,200]$ versus $\gamma$ for odd system sizes $L\in[7,17]$ in the presence of periodic boundary conditions. The vertical dashed line marks the onset of the chaotic regime identified from the study with HWBC (see Fig.~\ref{M-fig:CTDRevar} in main manuscript). Errors are indicated by symbol size.}
  \label{fig:CTDRevarPBC}
\end{figure}

For the homogeneous Fock state at unit density, the behaviour of the CTD observed for HWBC in Figs.~\ref{M-fig:CTDglobal} and \ref{M-fig:CTDRevar} of the main manuscript is equally found for the system in the presence of PBC, and the $\gamma$-regime of suppressed temporal fluctuations is qualitatively the same, as shown in Fig.~\ref{fig:CTDRevarPBC}.

\section{Chaotic phase from eigenvector properties}
\begin{figure}
  \centering
  \includegraphics[width=.9\columnwidth]{\figdir/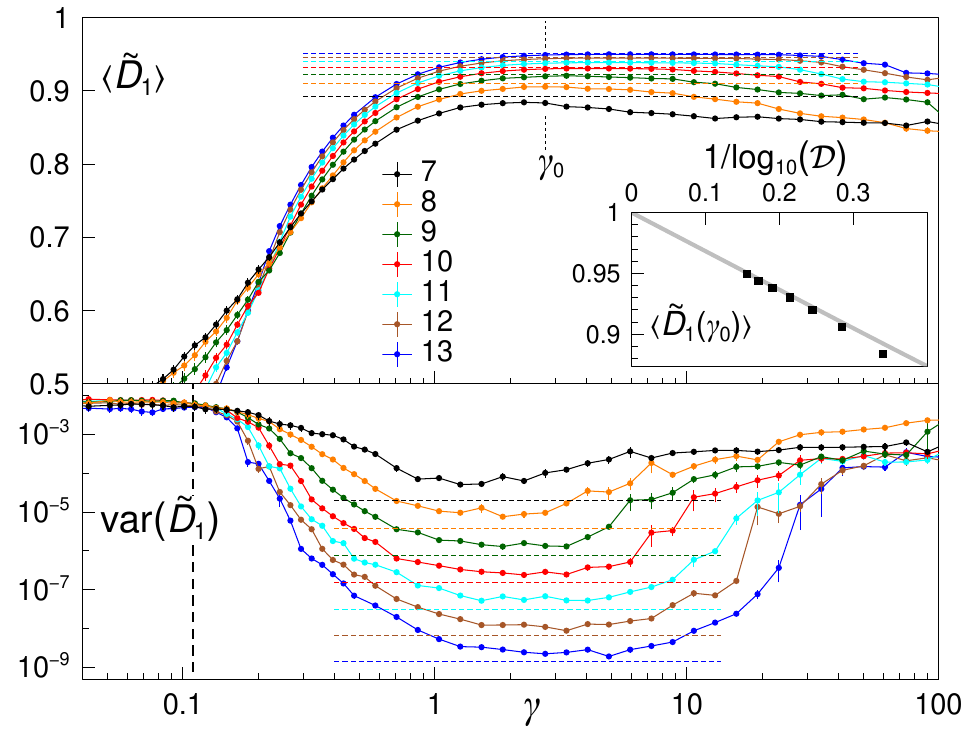}
  \caption{Mean (top) and variance (bottom) of $\tilde{D}_1$ from the 100 eigenstates closest in energy to $E=0$ (the energy of the initial state $\ket{\psi_0}$) as a function of $\gamma$ for varying system size $L\in[7,13]$ ($\mathcal{D}\in[868, 2\,600\,612]$). Horizontal dashed lines mark the corresponding values for GOE eigenvectors \cite{Pausch2020}. The inset shows $\langle \tilde{D}_1\rangle$ versus Hilbert space size $\mathcal{D}$ for $\gamma_0=2.74$, where the solid line is the GOE prediction. The vertical dashed line in the bottom panel marks the onset of the chaotic regime identified from the dynamical study (see Fig.~\ref{M-fig:CTDRevar} in main manuscript). When not visible, errors are contained within symbols size.}
  \label{fig:D1vsL}
\end{figure}

As demonstrated in Refs.~\cite{Pausch2020,Pausch2021,Pausch2022,PauschThesis}, the variance of the generalized fractal dimension $\tilde{D}_1 = -\sum_{\ket{\vb{n}}} |\braket{\vb{n}|\psi}|^2 \log |\braket{\vb{n}|\psi}|^2 /\log \mathcal{D}$ for close-in-energy eigenstates is a very sensitive probe of the emergence of quantum chaos. In Fig.~\ref{fig:D1vsL}, we show the identification of the system's chaotic phase from the behaviour of the mean and variance of $\tilde{D}_1$ for eigenstates close to the energy of the initial state $\ket{\psi_0}$ as a function of $\gamma$ and system size $L$.

The chaotic phase is revealed by the suppression of $\var(\tilde{D}_1)$ and the accompanying emergence of extended (ergodic) states  ($\tilde{D}_1\to 1$) in Fock space in the thermodynamic limit, in accordance with random matrix theory predictions [in this case for the Gaussian Orthogonal Ensemble (GOE)], as demonstrated in the inset to Fig.~\ref{fig:D1vsL} for $\gamma_0\equiv 2.74$, corresponding to the relative tunneling strength where the observed CTD temporal fluctuations achieve their minimum value as $L$ grows.

\section{Dynamics via TEBD for MPS}
\begin{figure}
  \centering
  \includegraphics[width=.95\columnwidth]{\figdir/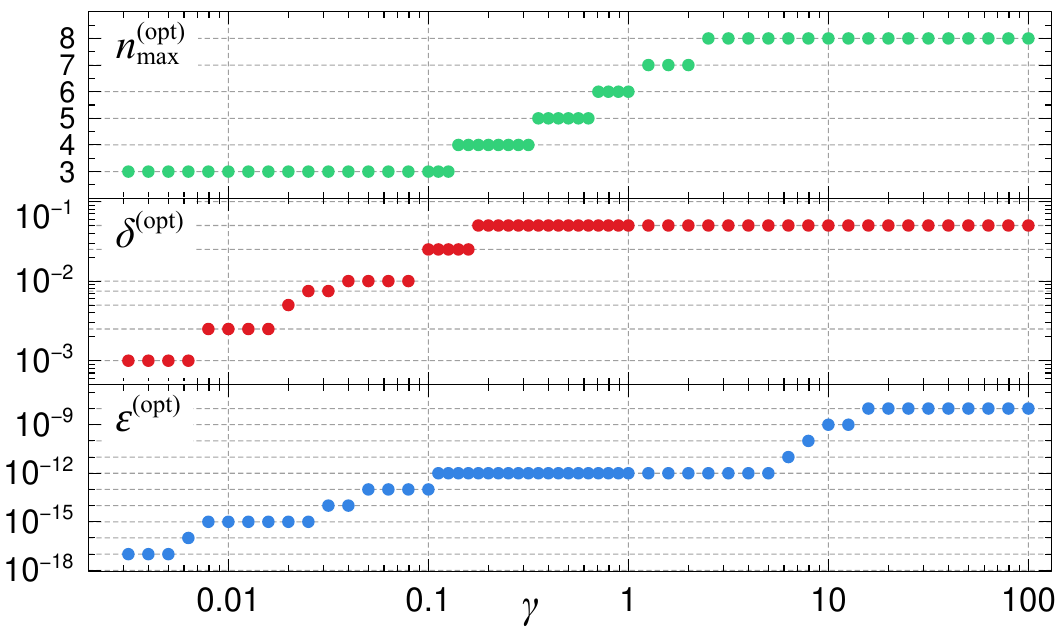}
  \caption{Optimal parameters used in   TEBD4-MPS simulations versus $\gamma$: (Top) Optimal maximum local occupation $n^{\rm (opt)}_{\rm max}$, (middle) optimal time step $\delta^{\rm (opt)}$, (bottom) optimal cutoff value $\varepsilon^{\rm (opt)}$.}
  \label{fig:MPS-Ladder}
\end{figure}

\begin{figure}
  \centering
  \includegraphics[width=.95\columnwidth]{\figdir/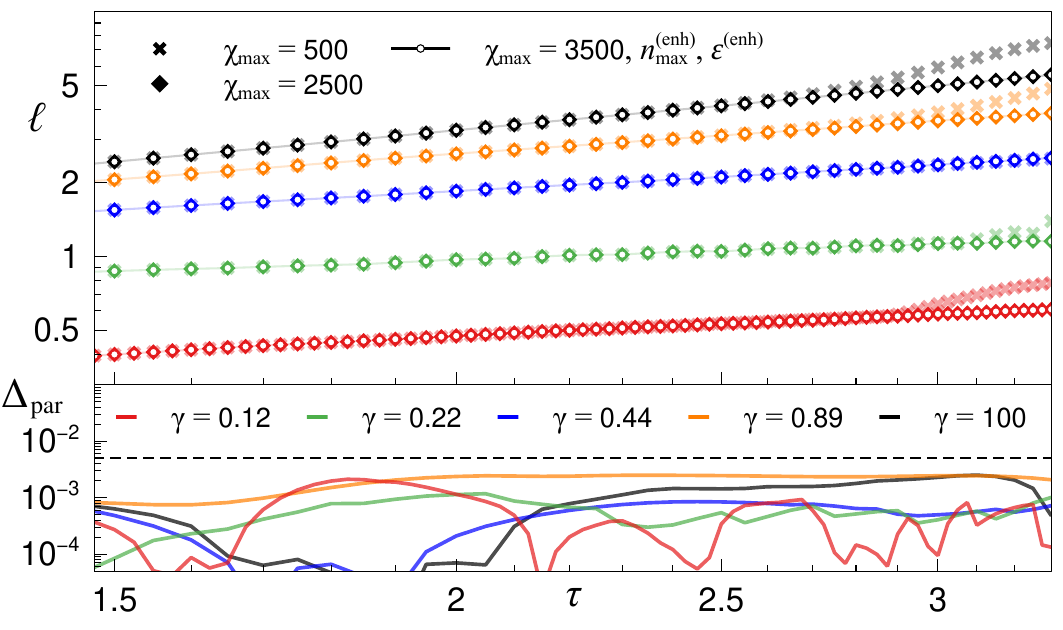}
  \caption{Comparison of signals for   various sets of TEBD4-MPS simulation parameters (see main text) for five $\gamma$ values (color coded). The top panel shows the CTD versus $\tau$ (log-log plot). The bottom panel displays the relative difference between the signal $\ell^{\rm (enh)}(\tau)$, computed using the enhanced set of parameters (see main text), and the optimal one. The horizontal dashed line marks the value $0.005$ used as a reference for convergence.}
  \label{fig:MPS-Enh-vs-Opt}
\end{figure}

We calculate the CTD for larger systems before the saturation regime via the fourth order time evolving block decimation (TEBD4) algorithm for matrix product states (MPS) \cite{Paeckel2019}. To perform the simulations, one needs to specify the value of a set of parameters: Maximum local occupation $n_{\rm max}$, time step $\delta$ and cutoff $\varepsilon$ for the MPS truncation after each time step. The corresponding optimal parameters are shown, as a function of $\gamma$, in Fig.~\ref{fig:MPS-Ladder}, and were chosen according to the convergence criteria discussed in the following, and with the aim of minimizing computational resources. In addition, one needs to set the maximum bond dimension $\chi_{\rm max}$ allowed for the MPS. We find that $\chi_{\rm max}=2500$ ensures a reliable signal up to $\tau=3.3$, as we demonstrate below and show in Fig.~\ref{fig:MPS-Enh-vs-Opt}.

To determine the optimal value of $n_{\rm max}$, we compare, up to $\tau=4$, the CTD signals for small systems and different values of $n_{\rm max}$ with the one corresponding to $n_{\rm max}=N$ (total boson number), i.e., with the exact dynamics, performed using the Chebyshev polynomial expansion described above. A signal is considered converged when the relative change with respect to the exact dynamics is at most $0.5\%$ at any time, i.e.,
\begin{equation}\label{eq:Conv-Occ}
 \textrm{max}_\tau{\left[ \frac{\left\lvert \ell_{N}(\tau)-\ell_{n_{\rm max}}(\tau) \right\rvert}{\ell_{N}(\tau)} \right]}\leqslant 0.005.
\end{equation}
The comparison is performed for different system sizes ($L=8,9,10\text{ and }15$) to check if these optimal values become size independent, which we find to be true. To verify the suitability of the obtained $n^{\rm (opt)}_{\rm max}$, we carry out the simulations for $L=40$ with $n_{\rm max}=n_{\rm max}^{\rm(opt)} + 1$, using now TEBD4, and compute the relative difference between this signal and the one corresponding to $n_{\rm max}^{\rm (opt)}$. For all values of $\gamma$ used in the pre-saturation analysis, this relative difference is below $0.5\%$. In the top panel of Fig.~\ref{fig:MPS-Ladder}, we show $n^{\rm (opt)}_{\rm max}$ as a function of $\gamma$, where $n^{\rm (opt)}_{\rm max}$ ranges from $3$ for $\gamma<0.14$ to $8$ for $\gamma>2$.

On the other hand, to determine the optimal value of the time step $\delta$, we exploit our knowledge about the short-time behaviour of the CTD \eqref{eq:CTD-Short-Time}. We do so by performing, for different values of $\delta$, the dynamics for the first five time steps (i.e., very short times) and $L=100$ (as the error associated with $\delta$ grows with system size), and by checking that the corresponding signal is converged with respect to the universal quadratic growth, according to
\begin{equation}
 \textrm{max}_\tau{\left\lbrace \frac{\left\lvert\log(4\tau^2)-\log{\left[ \ell_\delta(\tau) \right]}\right\rvert}{\log(4\tau^2)} \right\rbrace}\leqslant 0.005,
\end{equation}
where the comparison in log scale is appropriate since the signal tends to zero when $\delta$ is decreased. We have confirmed the suitability of the ensuing $\delta^{\rm (opt)}$  by computing, for six representative values of $\gamma$ ($0.012, 0.12, 0.22, 0.44, 0.89$ and $100$), the CTD signals for longer times with $\delta=\delta^{\rm (opt)}$ and $\delta=\delta^{\rm (opt)}/2$ and analysing the relative difference. For $\tau\in [2,3.3]$, the difference at any time is $\leqslant 0.8\%$ for $\gamma=0.22$, and below $0.2\%$ for all other gamma values. The middle panel of Fig.~\ref{fig:MPS-Ladder} displays $\delta^{\rm (opt)}$ as function of $\gamma$, ranging from $0.001$ for $\gamma<0.008$ to $0.05$ for $\gamma>0.16$.

Lastly, to determine the optimal value of the cutoff $\varepsilon$, we perform the dynamics up to $\tau=1$ for $L=100$ (the error associated with $\varepsilon$ also grows with system size) and for increasingly lower values of $\varepsilon$ (starting from $10^{-8}$), and compute the corresponding relative difference with respect to the signal with the reference cutoff $\varepsilon_{\rm min}$ ($10^{-18}$ for the lowest four $\gamma$ values and $10^{-16}$ for the rest). The convergence criterion reads
\begin{equation}
 \textrm{max}_\tau{\left[ \frac{\left\lvert \ell_{\varepsilon_{\rm min}}(\tau)-\ell_\varepsilon(\tau) \right\rvert}{\ell_{\varepsilon_{\rm min}}(\tau)} \right]}\leqslant 0.005.
\end{equation}
In the bottom panel of Fig.~\ref{fig:MPS-Ladder}, we show $\varepsilon^{\rm (opt)}$ as a function of $\gamma$, spanning the range from $10^{-17}$ for $\gamma<0.006$ to $10^{-8}$ for $\gamma>13$.

Now, in order to show the effects of a deficient $\chi_{\rm max}$ value, we performed the dynamics for seven $\gamma$ ($0.012, 0.050, 0.12, 0.22, 0.44, 0.89$ and $100$) with $\chi_{\rm max}=500$ (and all other parameters set to their optimal values). In the top panel of Fig.~\ref{fig:MPS-Enh-vs-Opt}, we compare the latter signals with the ones using $\chi_{\rm max}=2500$ (only five $\gamma$ values are shown for clarity). As one can see, deviations for the $\chi_{\rm max}=500$ signals are visible when approaching $\tau=3$.

To double check the validity of our optimal parameters and our choice of $\chi_{\rm max}$,
we simulate the dynamics for the same seven values of $\gamma$ and the enhanced set of parameters $\varepsilon^{\rm (enh)}=\varepsilon^{\rm (opt)}/100$, $n_{\rm max}^{\rm (enh)}=n_{\rm max}^{\rm (opt)}+1$, $\chi_{\rm max}=3500$. The comparison of the enhanced signal, $\ell^{\rm (enh)}(\tau)$, against the one for optimal parameters is shown  in Fig.~\ref{fig:MPS-Enh-vs-Opt} (only five $\gamma$ values are shown for clarity). We see that both signals are indistinguishable, independently of the value of $\gamma$, and the corresponding relative difference obeys $\Delta_{\rm par}(\tau)\equiv \left.\lvert\ell^{\rm (enh)}(\tau)-\ell^{\rm (opt)}(\tau)\rvert \middle/ \ell^{\rm (enh)}(\tau)\right. <0.003$ for all $\tau$ (bottom panel). Hence, we have demonstrated that our choice of parameters for the simulations ensures convergence of the CTD signal and the reliability of the analysis presented in the manuscript.

\subsection{Power-law growth and estimation of diffusion constant}

\begin{figure}
  \centering
  \includegraphics[width=.9\columnwidth]{\figdir/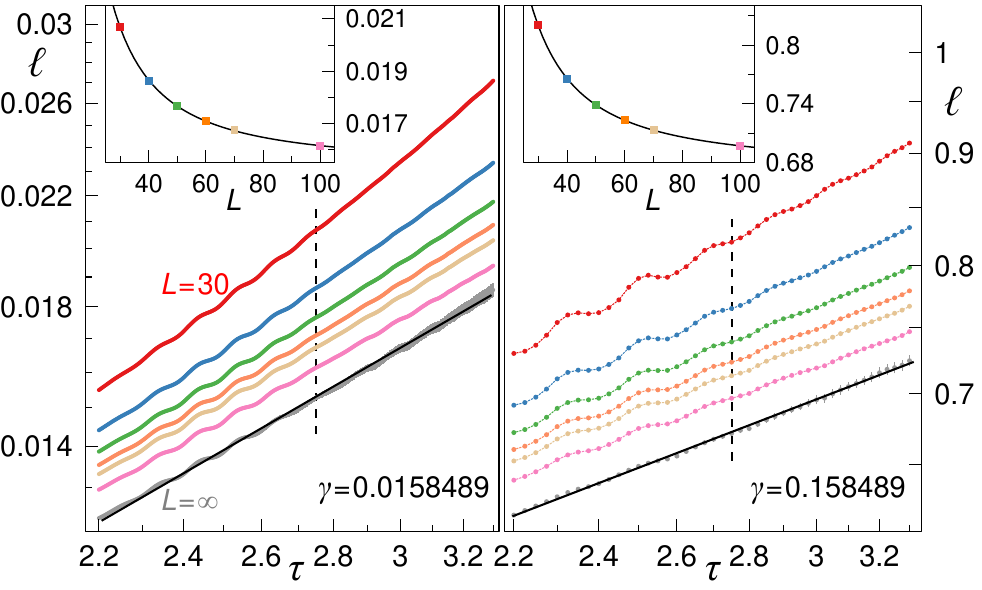}
  \caption{CTD versus time (log-log plot) for two values of $\gamma$ (as indicated) and varying system sizes $L\in\{30,40,50,60,70,100,\infty\}$ (from top to bottom). Straight black lines correspond to power-law guiding fits $\ell(\tau)=\alpha \tau^\beta$ for $L=\infty$ (see Fig.~\ref{M-fig:CTD-PLfits} in main manuscript). The data for $L=\infty$ ensue from the fits $\ell=a+b/L+c/L^2$ for each $\tau$ value (insets show the fits for $\tau=2.75$, highlighted by dashed lines in the main panels).}
  \label{fig:CTD-LinftyFits}
\end{figure}

One must note that, since our initial state is homogeneous in density, the presence of the system edges leaves an imprint on the CTD even at short times, and hence the signals for $\tau\in[2.2,3.3]$ still exhibit a dependence on $L$. The latter seems to die off as a polynomial in $L^{-1}$, which permits to extrapolate the CTD towards $L=\infty$ and in turn to characterize the corresponding power-law growth in this limit, as exemplary shown in Fig.~\ref{fig:CTD-LinftyFits}. The number of data points in the interval $\tau\in[2.2,3.3]$ ranges from 1101 for $\gamma<0.008$ to 23 for $\gamma>0.16$.

\begin{figure}
  \centering
  \includegraphics[width=.95\columnwidth]{\figdir/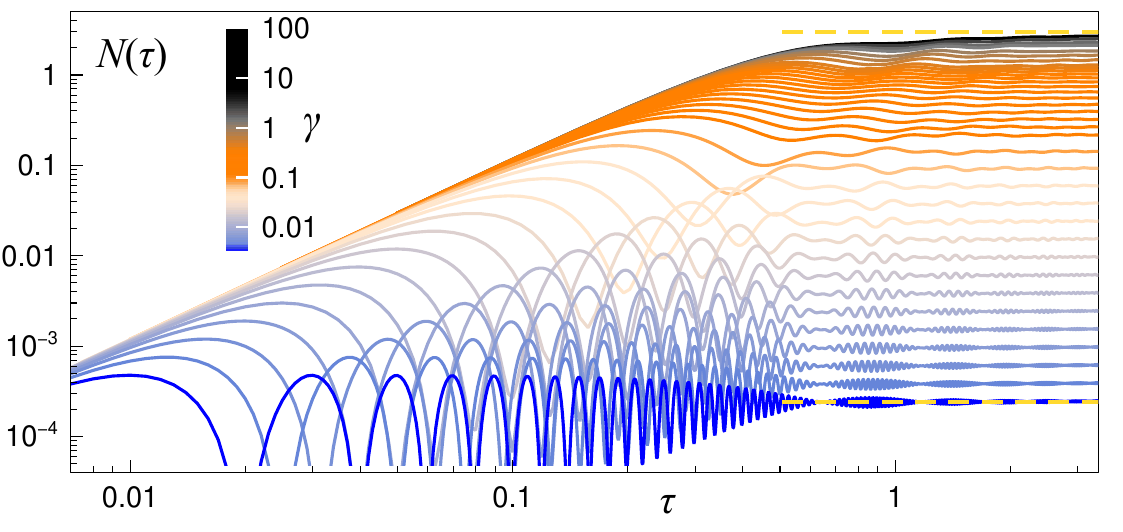}
  \caption{Time evolution of the norm $N(\tau)=\sum_{d\geqslant0}P(d,\tau)$ for varying $\gamma$ and $L=100$ obtained from TEBD4-MPS. Horizontal dashed lines mark the asymptotic value $3$ in the non-interacting limit [see Eq.~\eqref{eq:normP}] and the value $24\gamma^2$ for the lowest gamma considered ($\gamma=0.00316228$) according to Eq.~\eqref{eq:normgamma0}.}
  \label{fig:normP}
\end{figure}

A simple diffusive model for the random variable $x$, yields $\langle |x(\tau)| \rangle = \sqrt{4 D \tau/\pi}$. The interpretation of the CTD long-time increase as a diffusive process leads to the identification $\ell(\tau)/N(\tau)=\langle |x(\tau)| \rangle$, where $N(\tau)$ is the norm of the distribution $P(d,\tau)$ defined above [Eq.~\eqref{eq:Pd}]. In Fig.~\ref{fig:normP}, we demonstrate that the latter norm approaches a steady value for $\tau\gtrsim 2$ for any $\gamma$. Hence, from the CTD power-law fits $\ell(\tau)=\alpha \tau^\beta$ in the range $\tau\in[2.2,3.3]$, and the average value of the norm $\overline{N}$ in that interval, we estimate the diffusion coefficient as
\begin{equation}
 D=\frac{\pi}{4}\left(\frac{\alpha}{\overline{N}}\right)^2,
\end{equation}
in the $\gamma$-range where the diffusive growth of the CTD seems plausible (see Fig.~\ref{M-fig:CTD-PLfits}).

\begin{figure}
  \centering
  \includegraphics[width=.97\columnwidth]{\figdir/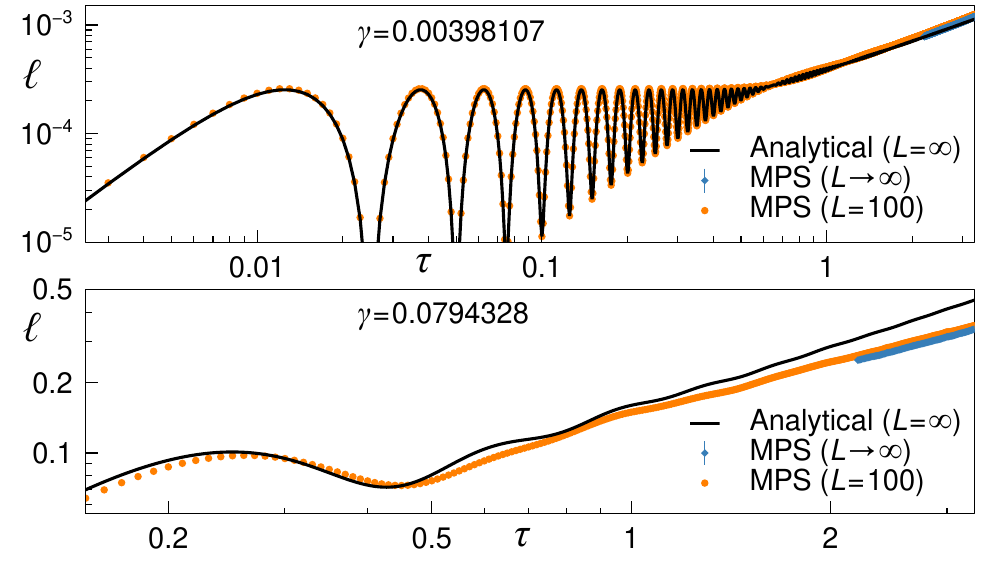}
  \caption{CTD versus time (log-log plot) for two values of $\gamma$ (as indicated) comparing the analytical result in the limit $\gamma\to0^+$, given by Eq.~\eqref{eq:CTDgamma0}, and numerics from TEBD4-MPS.}
  \label{fig:CTD-analyticalVSnumerics}
\end{figure}

Additionally, we check that the numerical CTD in the limit $\gamma\to0^+$ is very well described by Eq.~\eqref{eq:CTDgamma0}, as demonstrated in the upper panel of Fig.~\ref{fig:CTD-analyticalVSnumerics}. However, it must be emphasized, that already for $\gamma\approx 0.08$ ($U/J\approx 12$) the analytical description that follows from the fermionization approach \cite{Barmettler2012} fails to capture correctly the exponent that governs the steady growth of the CTD, as shown in the bottom panel of Fig.~\ref{fig:CTD-analyticalVSnumerics}. In fact,  Eq.~\eqref{eq:CTDgamma0}, if used for larger $\gamma$, can only give a ballistic spreading of correlations.

\end{document}